\documentclass{article}

\usepackage{graphicx}
\usepackage{amsfonts}

\title{Gravitational Tunnelling\\ of Relativistic Shells\footnote{\textit{To
appear in the Proceedings of the 6th International Symposium
on Frontiers in Fundamental and Computational Physics (FFP6),
September 26-29, Udine, ITALY}}}

\author{Stefano Ansoldi\footnote{Email: \texttt{ansoldi@trieste.infn.it}}%
{~}\footnote{Webpage: \texttt{http://www-dft.ts.infn.it/$\sim$ansoldi}}\\
{\small{}Dipartimento di Matematica e Informatica,}
{\small{}Universit\`{a} degli Studi di Udine,}\\
{\small{}and I.N.F.N. - Sezione di Trieste}\\
{\small{}via delle Scienze, 206 - I-33100 Udine (UD), ITALY}\\[3mm]
and\\[3mm]
Lorenzo Sindoni\footnote{Email: \texttt{potenzo17@yahoo.it}}\\
{\small{}Dipartimento di Fisica,}
{\small{}Universit\`{a} degli Studi di Trieste,}\\
{\small{}via A. Valerio, 2 - I-34127 Trieste (TS), ITALY}}

\begin{document}

\maketitle

\begin{abstract}
Thin shells in general relativity have been used in the past as keystones
to obtain realistic models of cosmological and astrophysical situations.
A crucial role for these developments was played by the compact description
of their dynamics in terms of Israel's junction conditions.
Starting from this geometrical formulation we present a problem related to
the WKB regime of shell dynamics and suggest a possible solution.
\end{abstract}

\begin{figure}
\begin{center}
\fbox{%
\vbox{%
\hbox{%
\parbox[b]{4.2cm}{\caption{\label{fig:conzer}Graphical representation of the apparent problem with the behavior
of the Euclidean momentum when one of the signs $\epsilon _{\pm}$ changes along the forbidden trajectory (here
$-$ is de Sitter and $+$ is Sch\-warz\-sch\-ild). It
seems that \emph{or} the momentum is vanishing at both extrema (top) \emph{or} it is continuous \dots{}}}%
{\hskip 2pt}%
\hbox{\includegraphics[width=8cm]{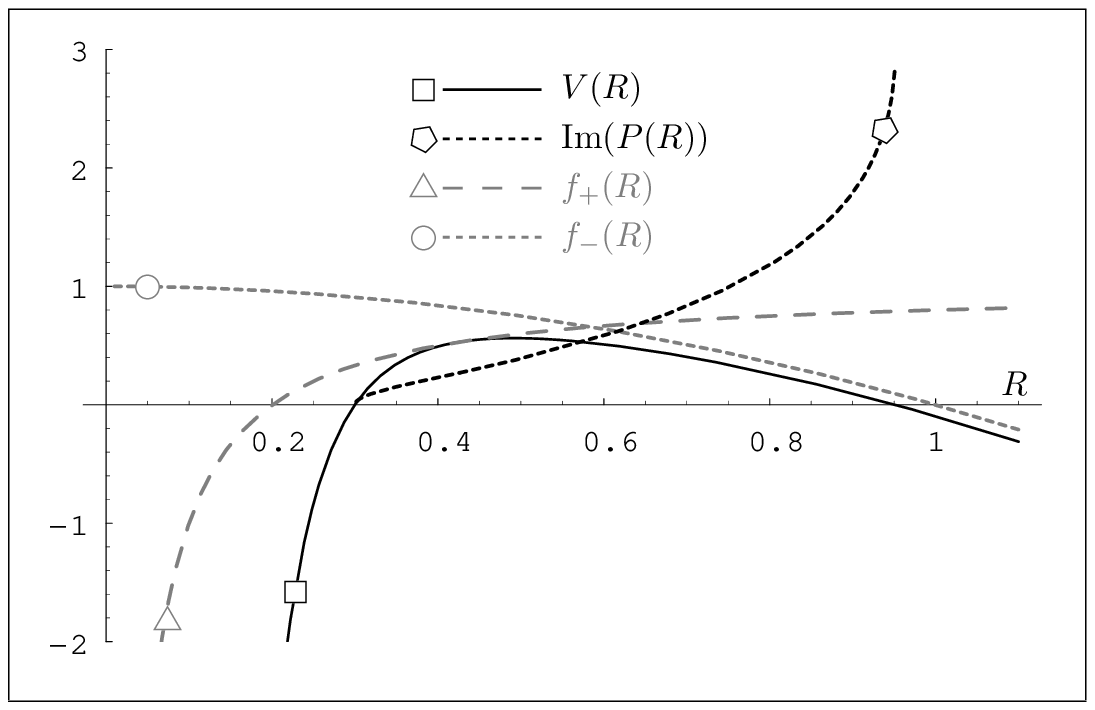}}%
}
\hbox{%
\hbox{\includegraphics[width=8cm]{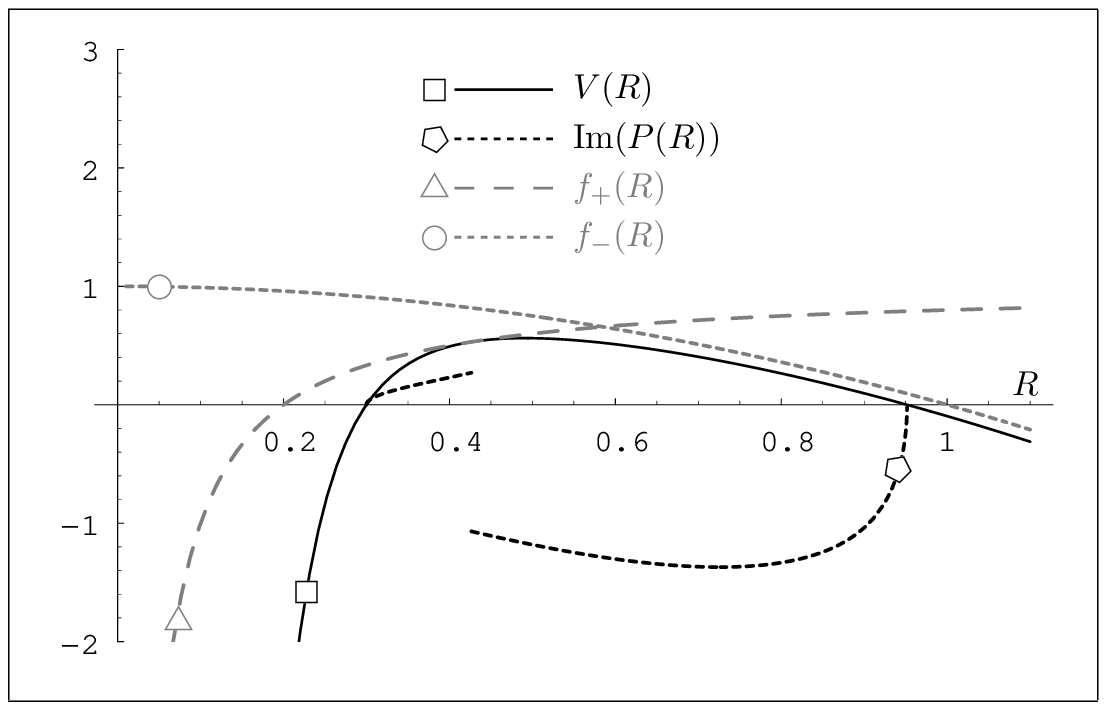}}%
{\hskip 2pt}%
\parbox[b]{4.2cm}{\dots{} (bottom). The two
behaviors correspond to the choice of different branches for the $\arctan$ function that emerges
from the $\tanh ^{-1}$ dependence in the conjugated momentum of equation ({\protect\ref{eq:effmom}}).
Although we are not discussing this problem here, this situation also requires a deeper analysis
in the Euclidean sector {\protect\cite{bib:hideki}}.\vspace{1mm}}%
}%
}}
\end{center}
\end{figure}

\begin{figure}
\begin{center}
\fbox{\hbox{\parbox[b]{6.2cm}{\caption{\label{fig:twimom}Representation of the Euclidean momentum
wrapped around a $(R,{\mathrm{Im}}(P))$-cone that replaces the standard $(R,{\mathrm{Im}}(P))$-plane
(Euclidean phase space).
At each value of $R$ the momentum takes values on an ${\mathbb{S}} ^{1}$ with radius proportional
to the current value of the radius $R$. In this way the momentum is \emph{both} continuous and
vanishing at the extrema. The gray straight segment in the picture corresponds to a segment
of the horizontal $R$ axis in figure {\protect\ref{fig:conzer}}.\vspace{1mm}}}%
{\hskip 2pt}\hbox{\includegraphics[width=6cm]{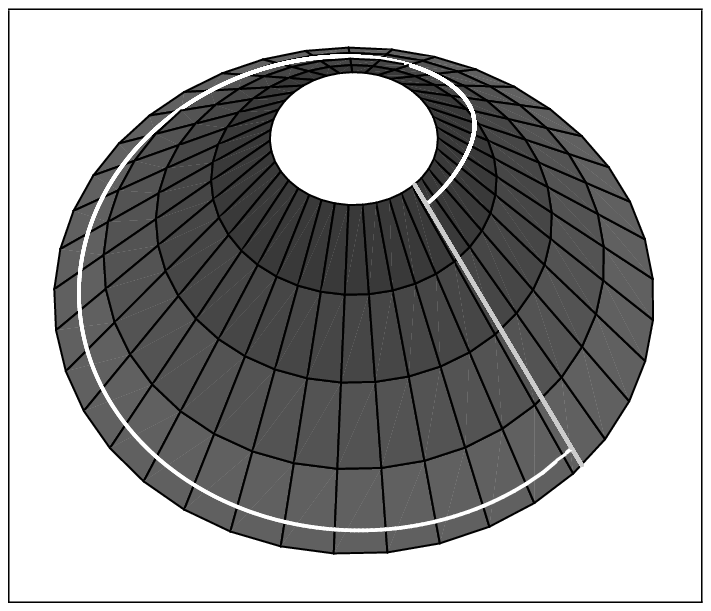}}}}
\end{center}
\end{figure}

General relativistic shells are an interesting system in general relativity
and because of the simple geometrical description of their dynamics provided by
Israel's junction conditions \cite{bib:NuCim1966.B44.....1I} they became
preferred models for many
crucial aspects of astrophysical and cosmological situations
(see \cite{bib:ClQuG2002..19..6321A}
for a more complete bibliography on the subject). Many of these models have
been developed under the assumption of spherical symmetry, but (as it
happens for instance in the case of gravitational collapse
\cite{bib:PhReL1965..14....57P})
this does not seem a severe restriction and it is likely that the obtained
results can
be extended to more general situations. On the other hand, the reduction in
the number of degrees of freedom that it is possible to obtain in the spherically
symmetric case makes simpler the development of effective models and more transparent
the discussion of the interesting subtleties that often appears in the geometrodynamics
of shells. Here we are, indeed, going to discuss one of these subtleties that already
manifests itself in the spherically symmetric case, where the junction conditions reduce
to just one equation\footnote{Following a
standard notation, quantities in the two spacetime regions separated by the shell are
identified by $\pm$ subscripts. We use square brackets ``$[\dots{}]$'' to denote their
jump in going from the ``$-$'' to the ``$+$'' side of the shell and an overdot,
``$\dot{\quad}$'', to indicate the derivative with respect to the proper time
measured by a shell-comoving observer.}
\begin{equation}
    [
        \epsilon ( \dot{R} ^{2} + f (R) ) ^{1/2}
    ]
    \equiv
    \epsilon _{-} ( \dot{R} ^{2} + f _{-} (R) ) ^{1/2}
    -
    \epsilon _{+} ( \dot{R} ^{2} + f _{+} (R) ) ^{1/2}
    =
    M(R)/R
    ,
\label{eq:sphjuncon}
\end{equation}
a first order integral of the second order equation of motion for the shell.
In (\ref{eq:sphjuncon}) $R$ is the radius of the shell (a function of the proper time $\tau$ of
an observer comoving with the shell); $M(R)$ describes the matter content of the shell
(i.e. it is related to its stress-energy tensor); $f _{\pm} (R)$ are the metric functions
in the two domains of spacetime separated by the shell when the line element is written
in the static form adapted to the spherical symmetry; $\epsilon _{\pm}$ are signs
(i.e. $0 , \pm 1$). Much of the discussion that follows is centered on these last
quantities, $\epsilon _{\pm}$, but, before embarking this program, we also remember that,
starting from an effective Lagrangian (the particular form of which is not our concern here),
we can compute the second order equation of motion that has (\ref{eq:sphjuncon})
as a first integral and also obtain the effective momentum \cite{bib:ClQuG1997..14..2727S}
conjugated to the only surviving degree of freedom $R$,
\begin{equation}
    P (R , \dot{R})
    =
    R
    \left[
        \tanh ^{-1}
            \left(
                \epsilon \dot{R} / ( \dot{R} ^{2} + f (R) ) ^{1/2}
            \right) ^{\mathrm{Sgn}(f(R))}
    \right ]
    .
\label{eq:effmom}
\end{equation}
Moreover, equation (\ref{eq:sphjuncon}) can be cast in the form of
an effective equation \cite{bib:PhReD1987..35..2961G,bib:ClQuG1997..14..2727S}
for the motion of a unitary mass particle with zero energy in a potential $V (R)$,
$\dot{R} ^{2} + V (R) = 0$.
Then all the solutions of (\ref{eq:sphjuncon}) are solutions of this effective equation
and \textit{viceversa}. This solves the problem of obtaining a qualitative
description of how the radius $R$ changes as a function of the proper time $\tau$. Of course
this is not the full story, since we still have to build up the
global structure of the spacetime in which the shell leaves. It is in this process that
we need also the information provided by the functions $f _{\pm} (R)$ and by the two signs
$\epsilon _{\pm}$. In particular when \textit{cutting and pasting} the Penrose diagrams
to build up the complete spacetime, $\epsilon _{\pm}$ select the sides of the Penrose
diagram crossed by the trajectory \cite{bib:PhReD1987..35..2961G}. Expressions for
$\epsilon _{\pm}$ can be obtained with little algebra,
$\epsilon _{\pm} = \mathrm{Sign} \left( M(R) \left( f _{-} - f _{+} \mp M ^{2} (R) / R ^{2} \right)\right)$,
and the points where $\epsilon _{\pm}$
change from $\pm 1$ to $\mp 1$ are the points in which $f _{\pm} (R)$ are tangent to $V (R)$, if
they exist. Since $f _{\pm} (R) \geq V (R)$ always, the signs can change i)
when the shell is crossing a region with $f _{\pm} (R) \leq 0$ or ii) \textit{along a classically
forbidden trajectory}\footnote{Our formulation here is far too synthetic and we refer the reader
to the literature on the subject (for example \cite{bib:PhReD1987..35..2961G,bib:PhReD1987..36..2919T})
for extended background material.}, where $V (R) > 0$. It is shown in \cite{bib:ClQuG1997..14..2727S}
that integrating the analytic continuation of (\ref{eq:effmom}) on the classically forbidden
trajectory we can compute WKB transition amplitudes for the
tunnelling process through the potential barrier; these amplitudes agree with those already
computed by other means\footnote{This is a strong argument in favor of an expression
for the effective momentum that, when evaluated along a classically forbidden trajectory,
differs from (\ref{eq:effmom}), also evaluated on a classically forbidden trajectory,
by a total derivative of a function of $R$, at most.} in \cite{bib:PhReD1980..21..3305L}.
In the cases discussed in \cite{bib:ClQuG1997..14..2727S} the signs $\epsilon _{\pm}$
are constant along the forbidden trajectory, but this is not always the case.
\textit{We are here interested in a more detailed analysis of those cases in which one of the
signs}, $\epsilon _{\pm}$, \textit{indeed changes}. Let us then see what happens to the momentum
$P (R , \dot{R})$. Since on a forbidden trajectory $V (R) > 0$, then $f (R) > 0$: we can thus forget the weird
exponent in (\ref{eq:effmom}).
Moreover from the effective equation we obtain that $\dot{R} ^{2} < 0$
i.e. $\dot{R}$ is purely imaginary and the momentum $P (R)$ also is purely imaginary, since
$\tanh ^{-1} (\imath \dots{}) = \imath \arctan (\dots{})$. Let us now assume there is an $\bar{R}$ along
the forbidden trajectory where, say, $\epsilon _{-}$ changes sign. This means that when $R \to \bar{R} ^{\pm}$
we have $(\dot{R} ^{2} + f _{-} (\bar{R})) ^{1/2} \to 0 ^{\pm}$ (or $0 ^{\mp}$) and
the argument of the $\arctan(\dots{})$ tends to $- \infty$ on one side and to $+ \infty$ on the
other. Correspondingly, choosing the standard branch of the multivalued function $\arctan(\dots{})$,
the Euclidean momentum has a discontinuity. We can try to cure this pathology by choosing
a different branch of $\arctan(\dots{})$: but then, following the evolution of the \emph{now continuous}
momentum till the second turning point, the offset introduced by the choice of the new branch makes
the momentum non-vanishing there; this seems again a difficult situation to accept. Apparently,
we thus face the unpleasant situation of i) having a discontinuous Euclidean momentum that vanishes at both turning
points \textbf{or} ii) having a continuous momentum that does not vanish at both turning points
(we incidentally point out that if we construct the Penrose diagrams associated to the two spacetimes
joined by the shell before and after the transition, some difficulties in their interpretation also
occur). This situation is pictured in figure \ref{fig:conzer} and now, after having stated the problem,
we proceed to propose a possible solution, by considering again our Euclidean momentum and following
its evolution from the first turning point.
It starts from zero and after some path on the $R$ line it reaches
$\bar{R}$. At this point we enforce its continuity and keep following it until the second turning
point, where we impose that \textbf{it is} zero. We said above that this cannot happen, but we
implicitly made an assumption, namely that \textit{the Euclidean momentum is a function
taking values in the real line}. Relaxing this assumption we are going to see that
not so much remains of the above problem. Figure \ref{fig:twimom} shows indeed that if we consider
the Euclidean momentum as a function that at each point $R$ along the forbidden trajectory takes
values in a circle (${\mathbb{S}} ^{1}$) of radius $R$, then we can make the momentum
both continuous and vanishing at both extrema! We end this contribution
referring the reader to \cite{bib:tocome} for an extended
discussion from the point of view of Euclidean quantum gravity.
\small{%

}
\end{document}